\documentclass[12pt]{article}
\usepackage{graphicx}
\begin{document}

\begin{center}
{\bf Power Laws and Gaussians for Stock Market Fluctuations}

\bigskip
\c{C}a\u{g}lar Tuncay and Dietrich Stauffer*

Department of Physics, Middle East Technical University

06531 Ankara, Turkey

caglart@metu.edu.tr \quad stauffer@thp.uni-koeln.de

\bigskip
* Visiting from Institute for Theoretical Physics, 

Cologne University, D-50923 K\"oln, Euroland
\end{center}

\textbf{Abstract}

The daily volume of transaction on the New York Stock Exchange and its
day-to-day fluctuations are analysed with respect to power-law tails as well 
long-term trends. We also model the transition to a Gaussian distribution
for longer time intervals, like months instead of days.

\textit{Keywords}: Fat tails, normal distribution, Pareto distribution.

PACS numbers: 89.65.Gh

\section{Introduction}
The statistical analysis of stock market fluctuations has a long tradition,
in economics \cite{lux} as well as physics \cite{gopi}. It is widely accepted
that the probability distribution function for relative price changes is 
neither a Gaussian nor a L\'evy distribution but has power-law or "fat" 
tails. Thus the probability $P_>(r)$ for the relative price change (more 
precisely for the change $r$ in the logarithm of the price) to be larger than
$r$ decays roughly as  
$$ P_>(r) \propto 1/r^3, \quad P(r) \propto 1/r^4 \eqno(1) $$
where $P(r)$ is the direct distribution, not the cumulative one.

Besides the price $x$ also the volume $V$ and the number of transactions $T$
(both per day) is of interest \cite{volume}, related through
$$ V = x T \quad . \eqno(2)$$
Here we are interested in daily closures and not in price 
changes during the trading day, and assume that one
stock is followed which does not split into different companies, merge with
other companies, or loses its identity for other reasons. For market indices
like the Dow Jones (DJIA), and changing its composition over the years, we 
simply {\it define} $T$ as the ratio $V/x$. Our $V$ is measured in dollars per 
day, or more generally in local currency units (lcu) per day.

In the next section we compare the long-time trends in $x, \, V$ and $T$ for 
the DJIA as well as for International Business Machines. Section 3 looks at 
the probability distribution functions for $V$ and $T$ and also for their
daily changes, analogously to the well-known return distributions of Eq.(1).
Section 4 offer a simple model to explain why the distribution of monthly 
returns deviates from Eq(1) and gets closer to a Gaussian \cite{kullmann,penna}.

\section{Long-time trends}

\begin{figure}[hbt]
\begin{center}
\includegraphics[scale=0.69]{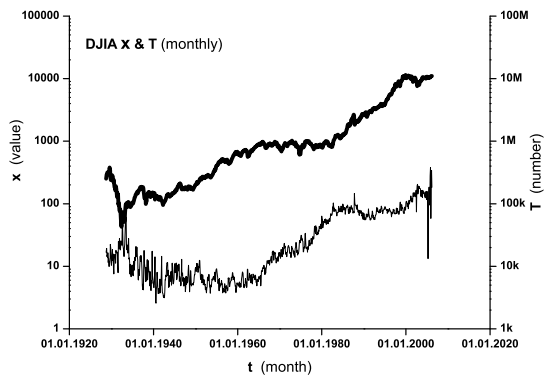}
\includegraphics[scale=0.69]{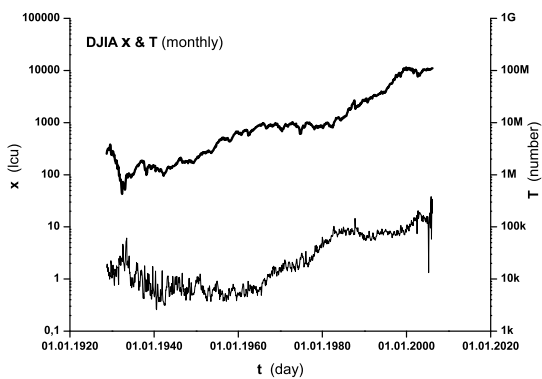}
\end{center}
\caption{
DJIA for $V$ (part a) and $T$ (part b) 
}
\end{figure}

\begin{figure}[hbt]
\begin{center}
\includegraphics[scale=0.69]{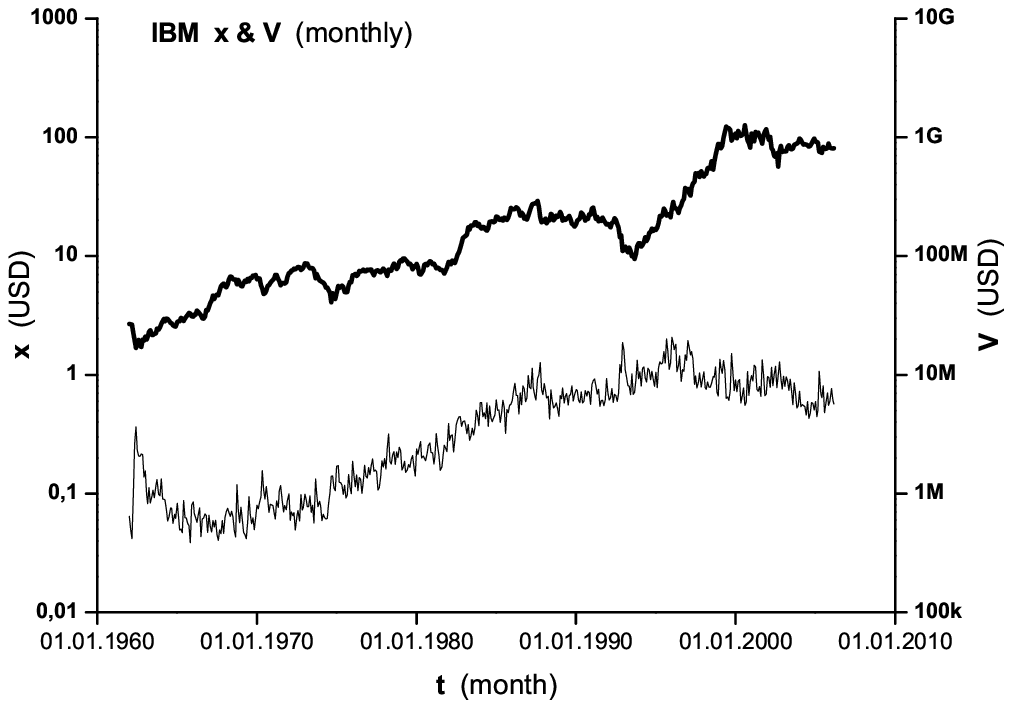}
\includegraphics[scale=0.69]{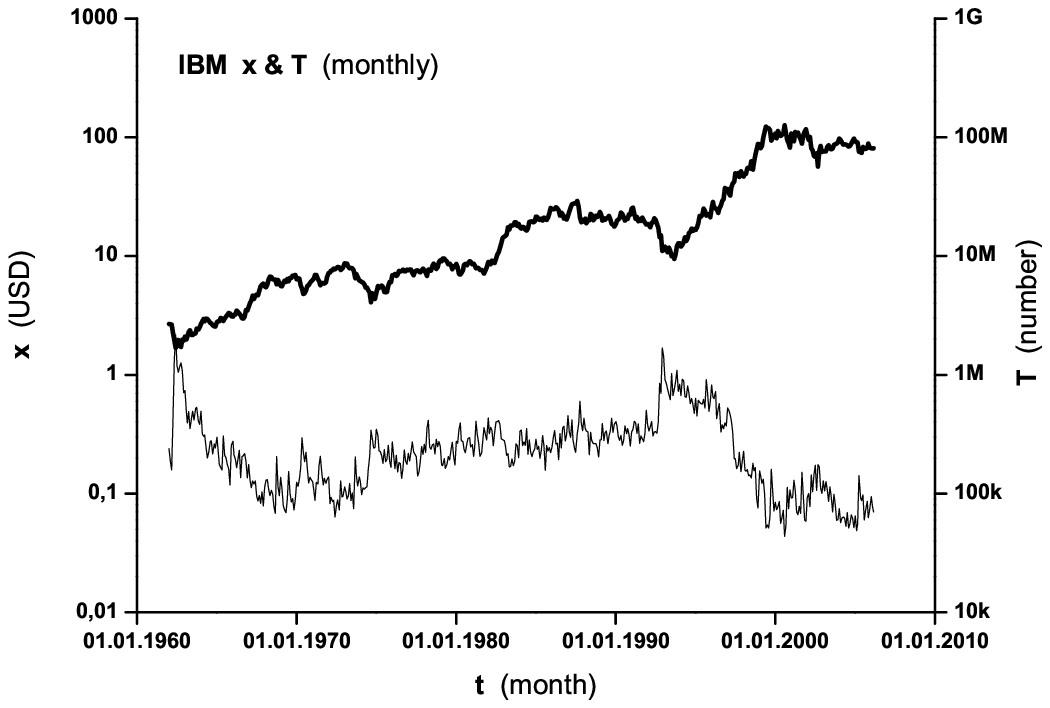}
\end{center}
\caption{
IBM for $V$ (part a) and $T$ (part b) 
}
\end{figure}

\begin{figure}[hbt]
\begin{center}
\includegraphics[scale=0.69]{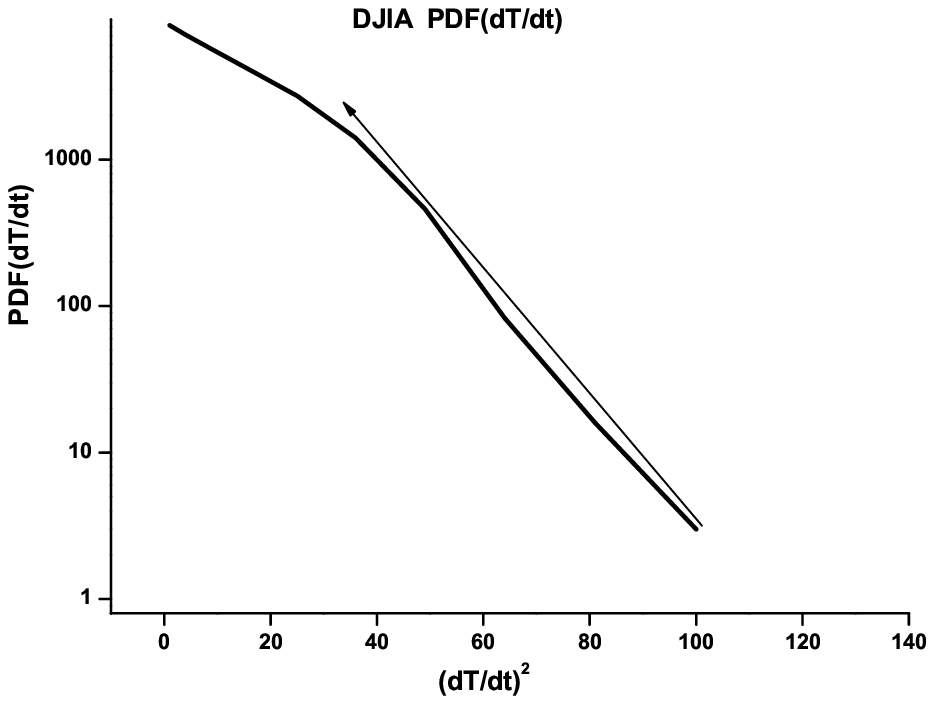}
\includegraphics[scale=0.69]{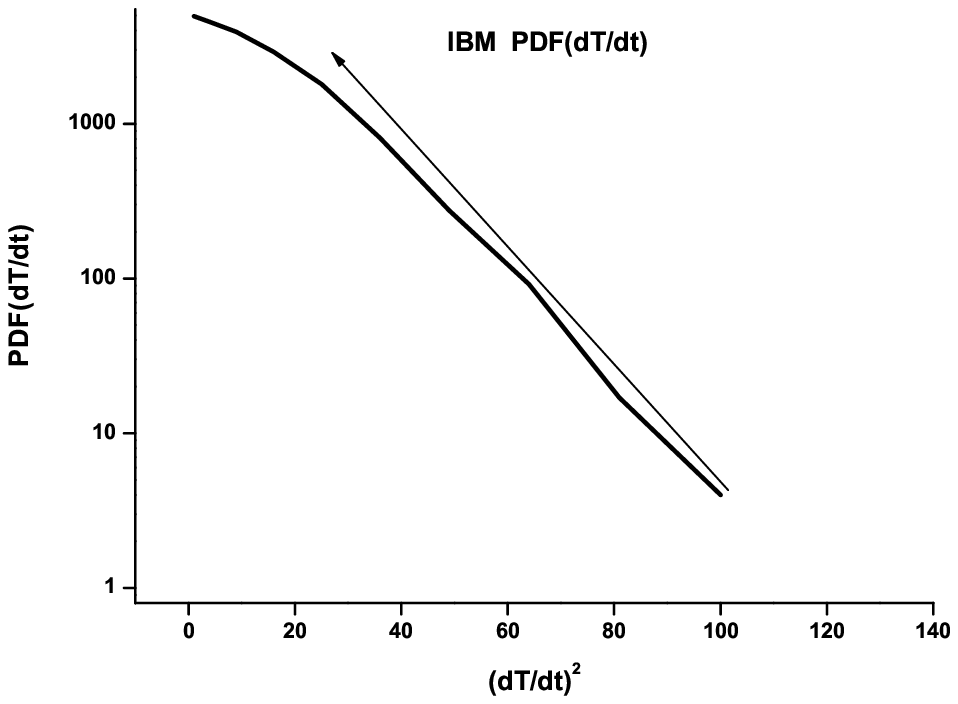}
\end{center}
\caption{
Probability distribution fnction for the daily changes in $T$,
plotted versus the squared change. The straight line corresponds to 
a Gaussian tail.
}
\end{figure}

\begin{figure}[hbt]
\begin{center}
\includegraphics[angle=-90,scale=0.5]{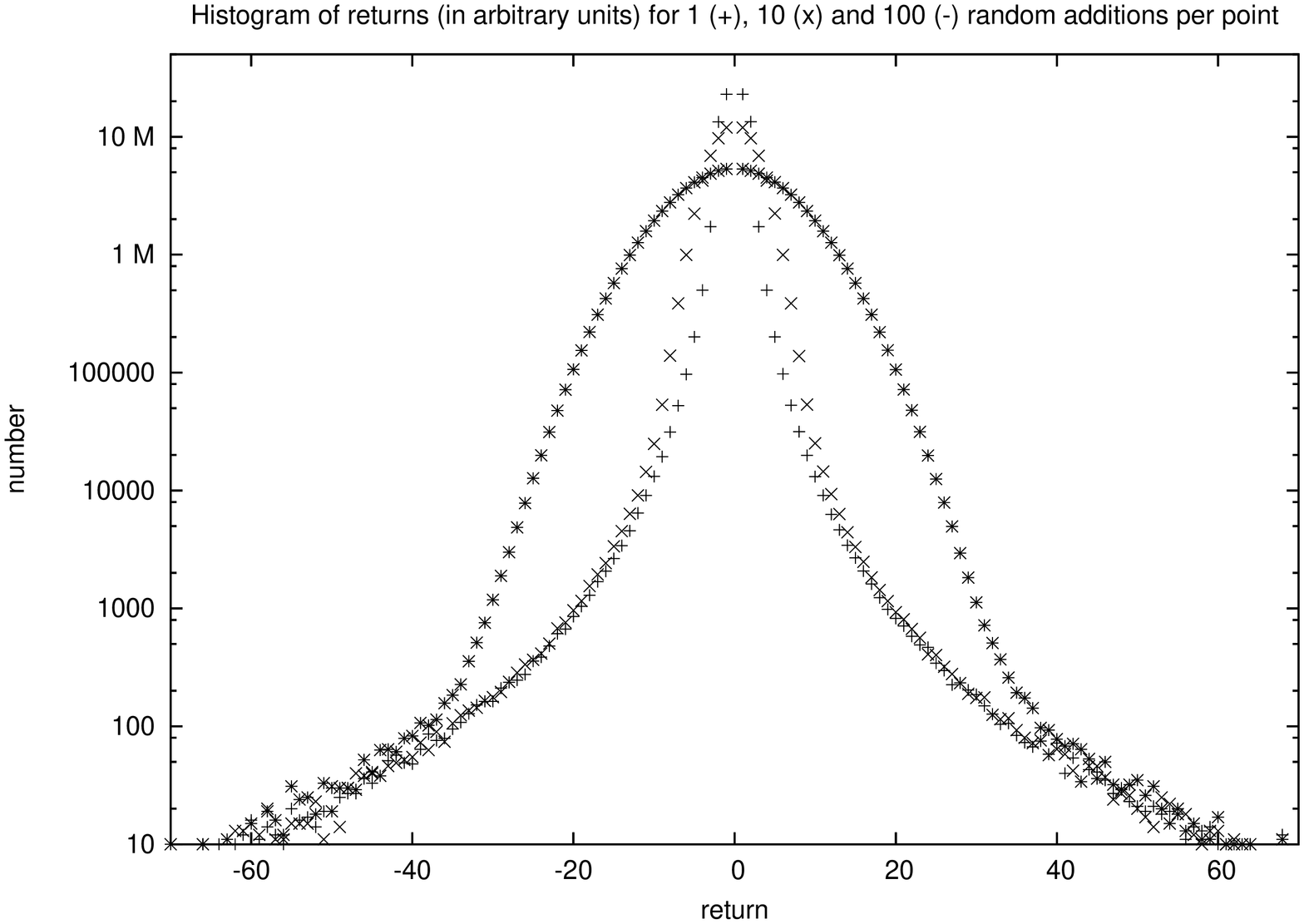}
\end{center}
\caption{
Crossover from power law for small $m$ to Gaussian distribution
for large $m$ in a superposition of Pareto power and random noise. 
$m = 1 (+), \; 10 (\times)$ and 100(stars).
}
\end{figure}

For the DJIA \cite{djia} we see in Fig.1 that the price $x(t)$ showed lot of
different behaviour since 1940 \cite{tuncay}, while the volume $V(t)$ increased
roughly exponentially. However, the fluctuations in $V$ were quite strong in 
1940 at the end of the Great Recession, and also in recent months. Perhaps 
these recent fluctuations signal a transition to a different regime, to be seen
in the coming years. The data for $T(t)$ show more coupling with the price $x$:
In the last half century, exponential increase in $x$ corresponded to a flat 
$T(t)$ while a flat $x$ was accompanied by an exponentially increasing $T$.

The exponential increase for $V(t)$ is seen for the daily volume, the weekly
volume, and the monthly volume, with the two latter quantities summing (not
averaging) over the trading days in that time interval. However, the large 
fluctuations in $V(t)$ in 1940 and now are seen clearer for the monthly 
than the weekly or daily volumes and thus are shown in Fig.1a,b.
Similar results (not shown) were obtained for the NASDAQ and S \& P 500
indices during the more recent decades. 

Analogous data for the single company IBM instead of the index DJIA are shown 
in Fig.2a,b, and here the above observations for $T$ cannot be repeated. Instead
a price increase twice corresponded to a falling $T$.

\section{Distributions}

The distributions of the daily volumes and transactions would not be
the analog of the return distributions
in Eq.(1) since the returns are the price {\it changes}. Thus we look in Fig.3
at the changes from one day to the next, $V(t)-V(t-1)$ and $T(t)-T(t-1)$
for DJIA and IBM. We see no good power laws; instead an exponential or 
Gaussian distribution fits the data better overall. This conclusion should 
be regarded as preliminary; it is possible that more accurate data over 
a larger number of decades \cite{gopi} would give a different result. 

Thus within the limits of our statistics the probability distribution functions
for volumes and their changes do not show the fat tails known from the 
price fluctuations. 

\section{Crossover}
 
The above discrepancy between price and volume fluctuations shows up if we
look at daily changes or even shorter times \cite{gopi,lux}. If instead
we look at price changes from one month to the other, a crossover towards
a more Gaussian distribution of returns $r$ is found \cite{kullmann}. We
now offer a simple model to explain this crossover from the power-law
of Eq.(1) to a Gaussian behavior,
$$ \ln[P(r)/P(0)] \propto - r^2 \quad .  \eqno(3)$$
We assume that the power law comes from intrinsic market behaviour, like
herding \cite{penna,kullmann}, while the Gaussian fluctuations come from
outside economic or political facts like inventions, wars, $\dots$ 
\cite{sornette}. For short times, few outside disturbances happen and
the power-law prevails; in longer time intervals more outside events 
influence the prices, and if these outside events are random, they 
accumulate to a Gaussian influence. 

As a quantitative model, we simulated a return distribution $P(r) \propto 
1/r^4$ with integer $r$; one unit in $r$ may correspond to a few tenths of a 
percent. (Each $r$ is rounded from $\pm z^{-1/3}$ where $z$ is taken randomly 
between 0 and 1.) Then each $r$ produced in this random way is modified by $m$ 
consecutive additions of $+1, \; -1$ and zero, with probabilities 1/4, 
1/4, and 1/2, respectively. The case $m=0$ is not shown and corresponds to 
Eq.(1); the cases $m = 1, \; 10, \; 100$ are shown in Fig.4 from $10^8$
samples. We see that $m=1$ barely differs from the power law, while for 
$m=10$ we see a Gaussian in the center and the power law in the tails. For
$m = 100$ nearly the whole range follows a Gaussian, and the fat tails are
so small that they would be visible only in high-quality statistics of real 
markets. Asymptotically, however, the tails should always follow the power law.

We thank Ondrej Hudak for comments on the manuscript.  DS thanks Emre Sururi 
Ta\c sci and Naz{\i}m Dugan for help with Turkish computers.

\end{document}